%% file: wiki-markup.tex
\documentclass[letterpaper]{sig-alternate-2013}

\usepackage{booktabs} 

\usepackage{graphicx}
\usepackage{mathtools}

\usepackage{amsmath,amssymb,amsfonts}

\usepackage{url}
\usepackage{verbatim}
\usepackage{multirow}
\usepackage{caption}
\usepackage{subcaption}
\usepackage[utf8]{inputenc}
\usepackage{arabtex}
\usepackage{algorithm}
\usepackage{algorithmicx}
\usepackage{algpseudocode}

\usepackage{algpseudocode}
\usepackage{mathtools}

\usepackage{ifpdf}
\usepackage{lmodern}
\usepackage{lipsum}
\usepackage[T1]{fontenc}
\usepackage{textcomp}
\usepackage{enumitem}
\usepackage{mdwlist}
\usepackage{pbox}
\usepackage{comment}
\usepackage{ifthen}
\usepackage{color}
\usepackage{colortbl,xcolor}
\usepackage{upgreek}
\usepackage{upgreek}
\usepackage{todonotes}
\usepackage{fixltx2e}
\MakeRobust{\overrightarrow}

\usepackage{bera}
\usepackage{listings,multicol}
\usepackage{xcolor}

\colorlet{punct}{red!60!black}
\definecolor{background}{HTML}{EEEEEE}
\definecolor{delim}{RGB}{20,105,176}
\colorlet{numb}{magenta!60!black}

\lstdefinelanguage{json}{
    basicstyle=\normalfont\ttfamily,
    numbers=left,
    numberstyle=\scriptsize,
    stepnumber=1,
    numbersep=8pt,
    showstringspaces=false,
    breaklines=true,
    frame=lines,
    tabsize=2,
    backgroundcolor=\color{background},
    literate=
     *{0}{{{\color{numb}0}}}{1}
      {1}{{{\color{numb}1}}}{1}
      {2}{{{\color{numb}2}}}{1}
      {3}{{{\color{numb}3}}}{1}
      {4}{{{\color{numb}4}}}{1}
      {5}{{{\color{numb}5}}}{1}
      {6}{{{\color{numb}6}}}{1}
      {7}{{{\color{numb}7}}}{1}
      {8}{{{\color{numb}8}}}{1}
      {9}{{{\color{numb}9}}}{1}
      {:}{{{\color{punct}{:}}}}{1}
      {,}{{{\color{punct}{,}}}}{1}
      {\{}{{{\color{delim}{\{}}}}{1}
      {\}}{{{\color{delim}{\}}}}}{1}
      {[}{{{\color{delim}{[}}}}{1}
      {]}{{{\color{delim}{]}}}}{1},
}

\usepackage{tikz}
\usepackage{longtable}

\usetikzlibrary{arrows}

\definecolor{Gray}{gray}{0.9}
\definecolor{LightCyan}{rgb}{0.88,1,1}

\specialcomment{notes}{\begingroup \color{blue}} { \endgroup}

\newcolumntype{!}{>{\global\let\currentrowstyle\relax}}
\newcolumntype{^}{>{\currentrowstyle}}

\newcommand{\si}{\begin{enumerate}}

\newcommand{\ei}{\end{enumerate}}

\makeatletter
\let\oldfootnote\footnote
\def\footnote{\@ifstar\footnote@star\footnote@nostar}
\def\footnote@star#1{{\let\thefootnote\relax\footnotetext{#1}}}
\def\footnote@nostar{\oldfootnote}
\makeatother

\DeclareGraphicsExtensions{.eps,.png}
\graphicspath{{figure/}{figure/eps/}{figure/png/}}

\newcommand{\superscript}[1]{\ensuremath{^{\textrm{#1}}}}

\newfont{\mycrnotice}{ptmr8t at 7pt}
\newfont{\myconfname}{ptmri8t at 7pt}
\let\crnotice\mycrnotice%
\let\confname\myconfname%

\begin{document}


\title{DAWT: Densely Annotated Wikipedia Texts across multiple languages}

\def\kloutinc{\superscript{*}}
\def\rutgers{\superscript{\dag}}

\numberofauthors{1} 
\author{
    \alignauthor Nemanja Spasojevic, Preeti Bhargava, Guoning Hu \\
    \affaddr{Lithium Technologies | Klout}\\
    \affaddr{San Francisco, CA}\\
    \email{\{nemanja.spasojevic, preeti.bhargava, guoning.hu\}@lithium.com}
} 

%
%


\maketitle

\begin{abstract}
\input{texfiles/wiki-markup_abstract}
\end{abstract}



\section{Introduction}
\label{section:introduction}
\input{texfiles/wiki-markup_introduction}

\section{Problem Statement}
\label{section:problem_statement}
\input{texfiles/wiki-markup_problem_statement}

\section{Contributions}
\label{section:contributions}
\input{texfiles/wiki-markup_contributions}

\section{Knowledge Base}
\label{section:generation}
\input{texfiles/wiki-markup_knowledgebase}

\section{DAWT Data Set Generation}
\label{section:generation}
\input{texfiles/wiki-markup_generation}

\section{Derived datasets}
\label{section:derived}
\input{texfiles/wiki-markup_results}

\section{Applications of the dataset}
\label{section:applications}
\input{texfiles/wiki-markup_applications}

\section{Accessing the dataset}
\label{section:access}
\input{texfiles/wiki-markup_access}

\section{Related Work}
\label{section:related}
\input{texfiles/wiki-markup_related}

\section{Conclusion and Future Work}
\label{section:conclusion}
\input{texfiles/wiki-markup_conclusion}




\bibliographystyle{abbrv}
\bibliography{bibliography}

\appendix
\input{texfiles/wiki-markup_appendix}

\end{document}

%% file: texfiles/wiki-markup_abstract.tex
In this work, we open up the DAWT dataset - Densely Annotated Wikipedia Texts across multiple languages. The annotations include labeled text mentions mapping to entities (represented by their Freebase machine ids) as well as the type of the entity.
The data set contains total of $13.6M$ articles, $5.0B$ tokens, $13.8M$ mention entity co-occurrences.
DAWT contains 4.8 times more anchor text to entity links than originally present in the Wikipedia markup. Moreover, it spans several languages including English, Spanish, Italian, German, French and Arabic. We also present the methodology used to generate the dataset which enriches Wikipedia markup in order to increase number of links. In addition to the main dataset, we open up several derived datasets including mention entity co-occurrence counts and entity embeddings, as well as mappings between Freebase ids and Wikidata item ids. 
We also discuss two applications of these datasets and hope that opening them up would prove useful for the Natural Language Processing and Information Retrieval communities, as well as facilitate multi-lingual research.

%% file: texfiles/wiki-markup_introduction.tex
Over the past decade, the amount of data available to enterprises has grown exponentially.
However, a majority of this data is unstructured or free-form text, also known as Dark Data\footnote{\url{https://en.wikipedia.org/wiki/Dark_data}}.
This data holds challenges for Natural Language Processing (NLP) and information retrieval (IR) tasks unless the text is
semantically labeled. Two NLP tasks that are particularly important to the IR community are:
\begin{itemize}[nolistsep,noitemsep]
  \item Named Entity Recognition (NER) - task of identifying an \emph{entity mention} within a text,
  \item Entity Disambiguation and Linking (EDL)  - task of linking the mention to its correct entity in a Knowledge Base (KB).
\end{itemize}
These tasks play a critical role in the construction of a high quality information network which can be further leveraged for a variety of IR and NLP tasks such as text categorization, topical interest and expertise modeling of users \cite{Spasojevic2016:experts, nemanja-lasta}. Moreover, when any new piece of information is extracted from text, it is necessary to know which real world entity this piece refers to. If the system makes an error here, it loses this piece of information and introduces noise. As a result, both these tasks require high quality labeled datasets with densely extracted mentions linking to their correct entities in a KB. 

Wikipedia has emerged as the most complete and widely used KB over the last decade.
As of today, it has around 5.3 million English articles and 38 million articles across all languages. 
In addition, due to its open nature and availability, Wikipedia Data Dumps have been adopted by academia and
industry as an extremely valuable data asset. Wikipedia precedes other OpenData projects like Freebase \cite{glorot2011domain} and DBpedia \cite{dbpedia} which were built on the foundation of Wikipedia.
The Freebase Knowledge Graph is the most exhaustive knowledge graph capturing 58 million entities and 3.17 billion facts.  The Wikipedia and Freebase data sets are large in terms of:
\begin{itemize}[nolistsep,noitemsep]
\item information comprehensiveness, 
\item wide language coverage, 
\item number of cross-article links, manually curated cross entity relations, and language independent entity identifiers.
\end{itemize}

Although these two data sets are readily available, Wikipedia link coverage is relatively sparse as only the first \emph{entity mention} is linked to the entity's Wikipedia article. This sparsity may significantly reduce the number of training samples one may derive from Wikipedia articles which, in turn, reduces the utility of the dataset. 
In this work, we primarily focus on creating the DAWT dataset that contains denser annotations across Wikipedia articles. We leverage Wikipedia and Freebase to build a large data set of annotated text where entities extracted from Wikipedia text are mapped to Freebase ids. Moreover, this data set spans multiple languages. 
In addition to the main dataset, we open up several derived datasets for mention occurrence counts, entity occurrence counts, mention entity co-occurrence counts and entity Word2Vec. We also discuss two applications of these datasets and hope that opening them up would prove useful for the NLP and IR communities as well as facilitate multi-lingual research.

%% file: texfiles/wiki-markup_problem_statement.tex

The Wikification problem was introduced by Mihalcea et al. \cite{Mihalcea:2007:WLD:1321440.1321475}, where task was to introduce hyperlinks to the correct wikipedia articles for a given mention. In Wikipedia, only the first mention is linked or annotated. 
In our task, we focus on densifying the annotations i.e. denser hyperlinks from mentions in Wikipedia articles to other Wikipedia articles.
The ultimate goal is to have high-precision hyperlinks with relatively high recall that could be further used as ground truth
for other NLP tasks.

For most of the supervised Machine Learning or NLP tasks, one of the challenges is gathering ground truth at scale.
In this work, we try to solve the problem of generating a labeled data set at large scale with the following constraints:

\begin{itemize}[noitemsep,nolistsep]
   \item The linked entity ids need to be unified across different languages. In Freebase, the machine id is same across different languages and hence, we annotate Wikipedia with Freebase machine ids,
   \item The dataset needs to be comprehensive (with large number of entities spanning multiple domains),
   \item The labels should be precise.
\end{itemize}

%% file: texfiles/wiki-markup_contributions.tex
Our contributions in this work are:
\begin{itemize}[noitemsep, nolistsep]
\item We extract a comprehensive inventory of mentions spanning several domains.
\item We densify the entity links in the Wikipedia documents by 4.8 times.
\item The DAWT dataset covers several more languages in addition to English such as Arabic, French, German, Italian, and Spanish.
\item Finally, we open up this dataset and several other derived datasets (such as mention occurrence counts, entity occurrence counts, mention entity co- occurrence counts, entity word2vec and mappings between Freebase ids and Wikidata item ids) for the benefit of the IR and NLP communities.
\end{itemize}

%% file: texfiles/wiki-markup_knowledgebase.tex
Our KB consists of about 1 million Freebase\footnote{Freebase was a standard community generated KB until June 2015 when Google deprecated it in favor of the commercially available Knowledge Graph API.} machine ids for entities.  
These were chosen from a subset of all Freebase entities that map to Wikipedia entities.
We  prefer to use Freebase as our KB since in Freebase, the same id represents a unique entity across multiple languages.
For a more general use, we have also provided the mapping from Freebase id to Wikipedia link and Wikidata item id (see Section \ref{secfreebasetowiki}).
Due to limited resources and usefulness of the entities, our KB contains approximately 1 million most important entities from among all the Freebase entities. 
This gives us a good balance between coverage and relevance of entities for processing common social media text.
To this end, we calculate an entity importance score \cite{Bhattacharyya-importance} using linear regression with features capturing popularity within Wikipedia links, and importance of the entity within Freebase. We used signals such as Wiki page rank, Wiki and Freebase incoming and outgoing links, and type descriptors within our KB etc. We use this score to rank the entities and retain only the top 1 million entities in our KB.

In addition to the KB entities, we also employ two special entities: \textbf{NIL} and \textbf{MISC}.
\textbf{NIL} entity indicates that there is no entity associated with the mention, eg. mention `the' within the sentence may link to entity \textbf{NIL}. This entity is useful especially when dealing with stop words and false positives. \textbf{MISC} indicates that the mention links to an entity which is outside the selected entity set in our KB. 

%% file: texfiles/wiki-markup_generation.tex

\begin{figure}[t]
\centering
\includegraphics[width=87mm]{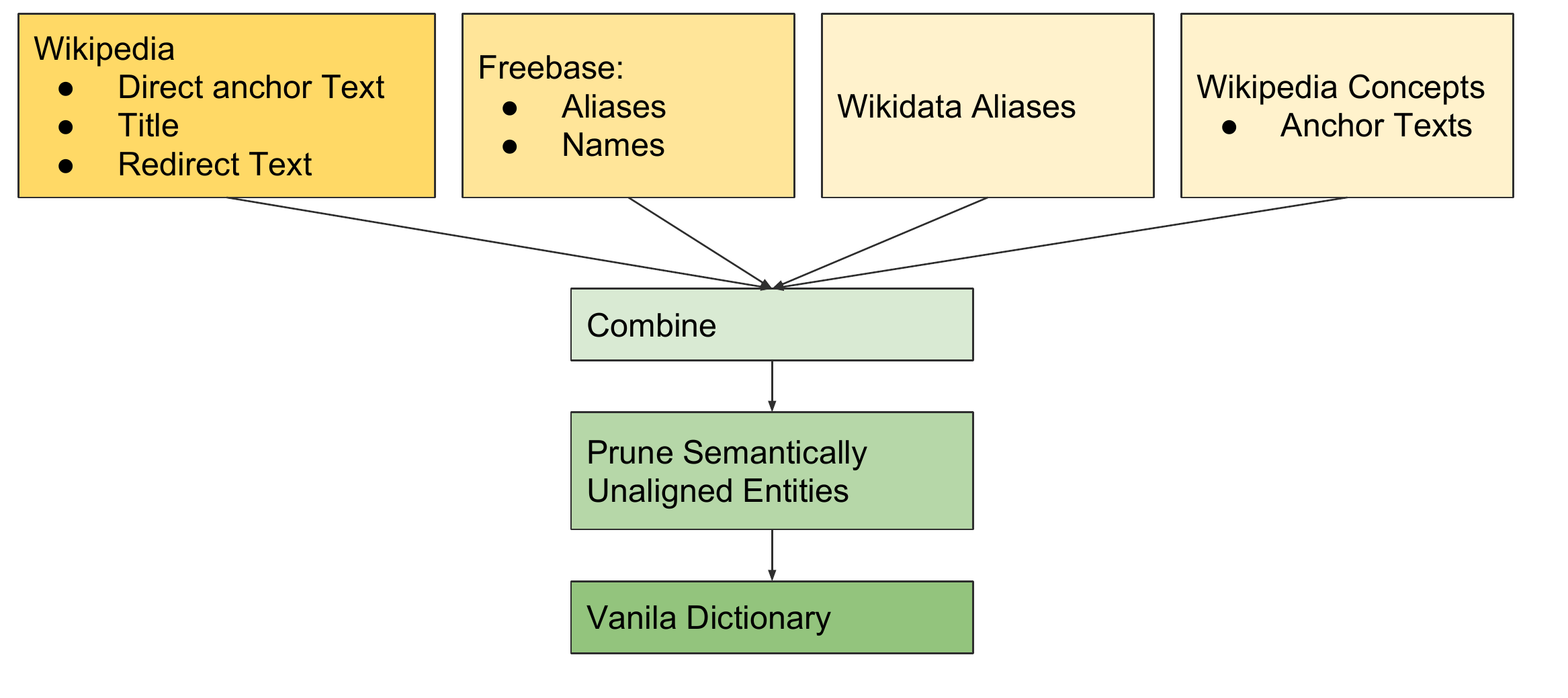}
\caption{Candidate Dictionary Generation Overview}
\label{fig:vanilla_dictionary}
\end{figure}

\begin{table*}[t]
\centering
\begin{tabular}{ |c|c|c|c|c|c|c| } \hline
\textbf{Language} & \textbf{Article} & \textbf{Unique Mention} & \textbf{Unique Entity} & \textbf{Unique Mention} & \textbf{Total Mention} & \textbf{Total CPU}\\
& &  \textbf{Count} &  \textbf{Count} & \textbf{Entity pairs} & \textbf{Entity links} & \textbf{time (days)}\ \\ \hline
en   &   5,303,722	& 5,786,727	  & 1,276,917	& 6,956,439	& 360,323,792  & 139.5\\ \hline
es   &   2,393,366	& 901,370	    & 224,695  	& 1,038,284	& 62,373,952  & 17.9\\ \hline
it   &   1,467,486	& 799,988	    & 211,687	  & 931,369	  & 47,659,715  & 14.2\\ \hline
fr   &   1,750,536	& 1,670,491	  & 423,603	  & 1,952,818	& 93,790,881 & 28.6  \\ \hline
de   &   1,818,649	& 2,168,723	  & 426,556	  & 2,438,583	& 103,738,278  & 20.2 \\ \hline
ar   &   889,007	  & 394,024	    & 186,787	  & 433,472	  & 12,387,715  & 1.6 \\ \hline
\end{tabular}
\caption{DAWT Data Set Statistics }
\label{table:data_set_stats}
\end{table*}

Our main goals when building the DAWT data set were to maintain high precision and increase linking coverage. 
As shown in Figure \ref{fig:vanilla_dictionary}, we first generate a list of candidate phrases mapping to the Wikipedia articles by combining:
\begin{itemize}[noitemsep,nolistsep]
  \item Wiki articles (direct anchor texts, titles of pages, redirect text to wiki pages)
  \item Freebase Aliases, and Freebase Also Known As fields related to entities
  \item Wikipedia Concepts (English anchor texts)
\end{itemize}
The initial candidate lists are pruned to remove outlier phrases that do not semantically align with the rest of the list.
As a semantic alignment metrics of two phrases, we used a combination of Jaccard similarity (both token and 3-gram character), edit distance (token and character), and largest common subsequence. We averaged the metrics and for each candidate in the list, we calculated average alignment against all the other candidates. As final step we remove all candidates, if any, with the lowest alignment scores.

For example, in the candidate set \{`USA', `US', `our', ... \} for entity \emph{USA}, phrase `our' does not align with rest of the cluster and is filtered out.
In addition to the candidate dictionary, we also calculate co-occurrence frequencies, based on direct links from Wikipedia article markup, between any 2 entities appearing within the same sentence.

To generate the DAWT dataset, we do the following. For each supported language, and for each Wiki page in the language:\begin{enumerate}[noitemsep,nolistsep]
  \item Iterate over the Wiki article and extract the set of directly linked entities.
  \item Calculate all probable co-occurring entities with the set of directly linked entities from Step 1.
  \item Iterate over the Wiki article and map all phrases to their set of candidate entities.
  \item Resolve phrases whose candidates have been directly linked from Step 1.
  \item For the remaining unresolved references, choose candidates, if any, with the highest probable co-occurrence with the directly linked entities.
\end{enumerate}
As a last step, the hyperlinks to Wikipedia articles in a specific language are replaced with links to their Freebase ids to adapt to our KB.

The densely annotated Wikipedia articles have on an average 4.8 times more links than the original articles. A detailed description of the data set, per  language, along with total CPU time taken for annotation is shown in Table \ref{table:data_set_stats}. All experiments were run on a 8-core 2.4GHz Xeon processor with 13 GB RAM. As evident, since English had the highest count of documents as well as entities and mentions, it took the maximum CPU time for annotation.

An example of the densely annotated text  in JSON format is given below:
\begin{lstlisting}[language=json,firstnumber=1,breaklines=true,caption=Annotated example in JSON file format]]
{
	"tokens": [{
		"raw_form": "Vlade"
	}, {
		"raw_form": "Divac"
	}, {
		"raw_form": "is"
	}, {
		"raw_form": "a"
	}, {
		"raw_form": "retired"
	}, {
		"raw_form": "Serbian"
	}, {
		"raw_form": "NBA"
	}, {
		"raw_form": "player",
		"break": ["SENTENCE"]
	}],
	"entities": [{
		"id_str": "01vpr3",
		"type": "PERSON",
		"start_position": 0,
		"end_position": 1,
		"raw_form": "Vlade Divac"
	}, {
		"id_str": "077qn",
		"type": "LOCATION",
		"start_position": 5,
		"end_position": 5,
		"raw_form": "Serbian"
	}, {
		"id_str": "05jvx",
		"type": "ORGANIZATION",
		"start_position": 6,
		"end_position": 6,
		"raw_form": "NBA"
	}],
	"id": "wiki_page_id:en:322505:01vpr3:Vlade_Divac"
}
\end{lstlisting}

%% file: texfiles/wiki-markup_results.tex

We also derive and open several other datasets from the DAWT dataset which we discuss here.

\input{texfiles/wiki-markup_results_frequnecy_counts}

\subsection{Freebase id to Wikidata id Mappings}\label{secfreebasetowiki}

In this data set, we use the Freebase machine id to represent an entity. To facilitate studies using Wikidata ids, which are also widely used entity ids in the literatures, we provide a data set that maps individual Freebase ids to Wikidata ids. This data set contains twice as many mappings as that from Google\footnote{\url{https://developers.google.com/freebase#freebase-wikidata-mappings}}. A summary comparison between these two mapping sets are shown in Table \ref{table:mappingComparison}, which lists the total numbers of mappings in 4 buckets:
\begin{itemize}[nolistsep,noitemsep]
    \item \textbf{Same}: A Freebase id maps to a same Wikidata id.
    \item \textbf{Different}: A Freebase id maps to different Wikidata ids.
    \item \textbf{DAWT Only}: A Freebase id only maps to a Wikidata id in DAWT.
    \item \textbf{Google Only}: A Freebase id only maps to a Wikidata id in Google.  
\end{itemize}
Note that the 24,638 different mappings are mainly caused by multiple Wikidata ids mapping to a same entity. For example, Freebase id 01159r maps to Q6110357 in DAWT and Q7355420 in Google, and both Q6110357 and Q7355420 represent 
the town Rockland in Wisconsin.

\begin{table}
\centering
\begin{tabular}{ |l|r| } 
\hline
Same & 2,048,531 \\ \hline
Different & 24,638 \\ \hline
DAWT Only & 2,362,077 \\ \hline
Google Only & 26,413 \\ \hline
\end{tabular}
\caption{Comparison of Freebase id to Wikidata id Mappings}
\label{table:mappingComparison}
\vspace{-0.1in}
\end{table}

\subsection{Entity Embeddings}
\label{sec:results_word2vec}
\input{texfiles/wiki-markup_results_word2vec}

%% file: texfiles/wiki-markup_results_frequnecy_counts.tex
%


\begin{table}

\begin{minipage}[b]{1\linewidth} 
\centering
\begin{tabular}{|c|c|}
\hline
\textbf{Mention}  & \textbf{Occurrence Count}   \\ \hline
Apple & 16104\\ \hline 
apple & 2742   \\ \hline 
Tesla & 822   \\ \hline 
\end{tabular}
\caption{Mention Occurrence Counts}
\label{table:mentioncounts}
\end{minipage}

\begin{minipage}[b]{1\linewidth} 
\centering
\begin{tabular}{|c|c|c|}
\hline
\textbf{Entity}  & \textbf{Occurrence Count}   \\ \hline
0k8z-Apple Inc & 39624 \\ \hline
014j1m-Apple (fruit) & 8727   \\ \hline  
05d1y-Nikola Tesla & 2777 \\ \hline
\end{tabular}
\caption{Entity Occurrence Counts}
\label{table:entitycounts}
\vspace{-0.1in}
\end{minipage}
\vspace{-0.1in}
\end{table}



\inputencoding{utf8}

\begin{table*}[t]
\centering
\begin{tabular}{ |c|c|c|c|c| } 
\hline
\textbf{Entity} & \textbf{Language} & \textbf{Surface Form} &
\textbf{Occurrence} & \textbf{Normalized} \\
& & & & \textbf{Occurrence} \\
\hline
\multirow{10}{*}{0k8z-Apple Inc} & \multirow{5}{*}{English} & Apple & 35166 & 71.97\% \\
& & Apple s & 4749	 & 9.72\% \\
& & Apple Inc & 2773 & 5.67\% \\
& & apple com & 2232 & 4.57\% \\
& & Apple Computer & 1534 & 3.14\% \\ 
\cline{2-5} 
& \multirow{5}{*}{French} &
Apple & 6305 & 85.97\% \\
& & apple & 466 & 6.35\% \\
& & d Apple	 & 228 & 3.11\% \\
& & Apple Inc & 109 & 1.49\% \\
& & Apple Computer & 96 & 1.31\% \\
\hline
\multirow{10}{*}{014j1m-Apple (fruit)} & \multirow{5}{*}{English} & apple & 4215 & 51.72\% \\
& & Apple & 1499 & 18.39\% \\
& & apples & 1409 & 17.29\% \\
& & Apples & 188 & 2.31 \% \\
& & Malus & 139 & 1.71 \% \\
\cline{2-5} 
& \multirow{5}{*}{French} & 
pomme & 2188 & 53.80\% \\
& & pommes & 1354 & 33.29\% \\
& & Pomme & 394 & 9.69\% \\
& & Pommes & 81 & 1.99\% \\
& & pommeraie & 13 & 0.32\% \\
\hline  
\multirow{10}{*}{05d1y-Nikola Tesla} & \multirow{5}{*}{English} & Tesla & 2391 & 66.79 \% \\
& & Nikola Tesla & 1043 & 29.13\% \\
& & Nikola Tesla s	 & 71 & 1.98\% \\
& & Tesla Nikola & 28 & 0.78\% \\
& & Nicola Tesla & 13 & 0.36\% \\ 
\cline{2-5} 
& \multirow{5}{*}{French} &
Tesla & 587 & 67.94\% \\
& & Nikola Tesla & 270 & 31.25\% \\
& & Nicolas Tesla & 5 & 0.58\% \\
& & Nicola Tesla & 1 & 0.11\% \\
& & nikola tesla & 1 & 0.11\% \\
\hline  
\multirow{8}{*}{0dr90d-Tesla Motors} & \multirow{5}{*}{English} & Tesla &1559 & 74.63\% \\
& & Tesla Motors & 437 & 20.92\% \\
& & Tesla Roadster & 22 & 1.05\% \\
& & teslamotors & 11 & 0.53\% \\
& & Tesla Motors Inc & 9 & 0.43\% \\
\cline{2-5} 
& \multirow{3}{*}{French} & 
Tesla & 110 & 68.32\% \\
& & Tesla Motors & 49 & 30.44\% \\
& & voitures Tesla & 2 & 1.24\% \\
\hline  
\end{tabular}
\caption{Entity surface form variation across languages (English and French)}
\label{table:entity_examples}
\vspace{-0.1in}
\end{table*}

\begin{table*}[t]
\centering
\begin{tabular}{|c|c|c|c|}
\hline
\textbf{Mention} & \textbf{Entity}  & \textbf{Co-occurrence}& \textbf{Normalized}  \\
& & & \textbf{Co-occurrence}   \\ \hline
\multirow{5}{*}{Apple} & 0k8z-Apple Inc. & 6738 & 87.8\% \\  
& 014j1m-Apple (fruit) & 422 & 5.5\%  \\ 
& 019n\_t-Apple Records & 302 & 3.9\%  \\  
& 02hwrl-Apple Store & 87 & 1.1\%  \\ 
& 02\_7z\_-Apple Corps. & 84 & 1.1\%   \\ \hline 
\multirow{5}{*}{apple} & 014j1m-Apple & 1295 & 85.3\% \\
& 01qd72-Malus & 157 & 10.3\% \\ 
& 02bjnm-Apple juice & 46 & 3.0\% \\ 
& 0gjjvk-The Apple Tree & 15 & 1.0\% \\
& 0k8z-Apple Inc. & 1 & 0.1\% \\ \hline
\multirow{5}{*}{Tesla} & 05d1y-Nikola Tesla & 327 & 49.5\% \\  
& 0dr90d-Tesla Motors & 162 & 24.5\%  \\ 
& 036wfx-Tesla (Band) & 92 & 13.9\%  \\ 
& 02rx3cy-Tesla (Microarchitecture) & 38 & 5.7\% \\ 
& 03rhvb-Tesla (Unit) & 29 & 4.4\%  \\ \hline 
\end{tabular}
\caption{Mention Entity Co-occurrences}
\label{table:mentionEntityCoCounts}
\vspace{-0.1in}
\end{table*}

\subsection{Mention Occurrences}\label{subsec:MentionOccur}
This dataset includes the raw occurrence counts for a mention $M_{i}$ in our corpus and KB. Table \ref{table:mentioncounts} shows the raw counts for mentions ``Apple", ``apple" and ``Tesla".

\subsection{Entity Occurrences}\label{subsec:EntityOccur}
This dataset includes the raw occurrence counts for an entity $E_{j}$ in our corpus and KB. Table \ref{table:entitycounts} shows the raw counts for entities \emph{Apple Inc.}, \emph{Apple (fruit)} and \emph{Nikola Tesla}. 
We also generate separate dictionaries for each language. Table \ref{table:entity_examples} shows the different surface form variations and occurrence counts of the same entity across different languages.

\subsection{Mention To Entity Co-occurrences}\label{subsec:MentionEntityOccur}

This dataset includes the co-occurrence counts of mentions and entities. This is particularly useful for estimating the prior probability of a mention $M_{i}$ referring to a candidate entity $E_{j}$ with respect to our KB and corpora. 
Table \ref{table:mentionEntityCoCounts} shows the raw and normalized mention entity co-occurrences for the mentions ``Apple" and ``apple" and different candidate entities. As evident, the probability of mention ``Apple" referring to the entity \emph{Apple Inc.} is higher than to the entity \emph{Apple (fruit)}. However, ``apple" most likely refers to the entity \emph{Apple (fruit)}. Similarly, the mention ``Tesla" most likely refers to the entity \emph{Nikola Tesla}. 


%% file: texfiles/wiki-markup_results_word2vec.tex
There have been many efforts on learning word embeddings, i.e., vector space representations of words \cite{deerwester1990indexing, mikolov2013efficient, Mikolov2013word2vec}. Such representations are very useful in many tasks, such as word analogy, word similarity, and named entity recognition. Recently, Pennington et al. \cite{glove2014} proposed a model, GloVe, which learns word embeddings with both global matrix factorization and local context windowing. They showed that obtained embeddings captured rich semantic information and performed well in the aforementioned tasks. 

There are several word-vector data sets available on GloVe's website\footnote{ \url{http://nlp.stanford.edu/projects/glove/}}. However, they only contain embeddings of individual words and thus have several limitations:
\begin{itemize}[nolistsep,noitemsep]
	\item Language dependent
	\item Missing entities that cannot be represented by a single word
	\item May not properly represent ambiguous words, such as "apple", which can be either the fruit or the technology company. 
\end{itemize}
To facilitate research in this direction, we provide an entity-embedding data set that overcomes the above limitations. This date set contains embeddings of Wiki entities with 3 different vector sizes: 50, 300, and 1000. They were generated with the GloVe model via the following steps:
\begin{enumerate}[nolistsep,noitemsep]
  \item Represent each Wiki document across all languages as a list of entities: There are about 2.2B total entities and 1.8M unique entities in these documents. 
  \item Use the open source GloVe code\footnote{\url{https://github.com/stanfordnlp/GloVe}} to process these documents: For each vector size, we ran 300 iterations on a GPU box with 24 cores and 60G dedicated memory. Other runtime configurations were the same as default. In particular, we:
  \begin{itemize}[nolistsep,noitemsep]
    \item Truncate entities with total count < 5
    \item Set window size to be 15 
  \end{itemize}
Execution time was roughly proportional to the vector size. It took about 25 minutes to run 1 iteration when size is 1000. Among the 1.8 M unique entities, the GloVe model were able to generate embeddings for about 1.6 M entities.
\end{enumerate}

To evaluate these embeddings, we compared them with one of the GloVe word embeddings, which was also generated from Wikipedia data\footnote{The data is available at \url{http://nlp.stanford.edu/data/glove.6B.zip}}, on the word/entity analogy task. This task is commonly used to evaluate embeddings \cite{mikolov2013efficient, Mikolov2013word2vec, glove2014} by answering the following question: \textit{Given word/entity X, Y, and Z, what is the
word/entity that is similar to Z in the same sense as Y is similar to X?} For example, given word "Athens", "Greece", and "Paris", the right answer is "France".

Here we used the test data provided by \cite{mikolov2013efficient}\footnote{The data set is available at \url{http://www.fit.vutbr.cz/~imikolov/rnnlm/word-test.v1.txt}}. This test set contains 5 semantic and 9 syntactic relation types. For each word in this data set, we find the corresponding Freebase id using the mapping between Freebase ids and english Wikidata urls. Thus, we obtain a test set that contains relations between entities. Note that when we could not find a Freebase id for a word, all the associated relations were removed from the test set.

We then ran the test on the 5 semantic types. Syntactic relations were excluded from this test because most of the time the task is trivial when one can correctly link words to entities. For example, when both "bird" and "birds" are linked to entity \emph{015p6-Bird}, and "cat" and "cats" are linked to entity \emph{01yrx-Cat}, the analogy among them is obvious without examining the underlining embeddings.

Table \ref{table:wordAnalogyComparison} shows the accuracy (in \%) obtained from our entity embeddings with vector sizes of 50, 300, and 1000. In comparison, it also shows the accuracy from GloVe word embeddings with vector sizes of 50, 100, 200, and 300. Entity embeddings have better performance with vector size of 50. As we increase vector size, word embeddings perform significantly better and outperform entity embeddings when the vector size is 200 or higher. The degraded performance of entity embeddings may due to less training data, since our entity embeddings were obtained from 2.2B tokens, where GloVe's word embeddings were obtained from 6B tokens.   

\begin{table*}[t]
\centering
\begin{tabular}{ |l|c|c|c|c|c|c|c| } 
\hline
\multirow{2}{*}{\textbf{Relation}} & \multicolumn{4}{c}{\textbf{GloVe Word dimensionality}} & \multicolumn{3}{|c|}{\textbf{DAWT Entity dimensionality}} \\
\cline{2-8}
& 50 & 100 & 200 & 300 & 50 & 300 & 1000 \\
\hline
Capital-World & 74.43 & 92.77 & 97.05 & 97.94 & 93.24 & 93.95 & 91.81 \\
\hline
City-in-State & 23.22 & 40.10 & 63.90 & 72.59 & 68.39 & 88.98 & 87.90 \\
\hline
Capital-Common-Countries & 80.04 & 95.06 & 96.64 & 97.23 & 78.66 & 79.64 & 71.54 \\
\hline
Currency & 17.29 & 30.05 & 37.77 & 35.90 & 43.88 & 13.56 & 2.93 \\
\hline
Family & 71.05 & 85.09 & 89.18 & 91.23 & 66.96 & 72.51 & 75.15 \\
\hline
Average & 53.21 & 68.61 & 76.91 & 78.98 & 70.23 & 69.73 & 65.87 \\
\hline  
\end{tabular}
\caption{Accuracy of Semantic Analogy}
\label{table:wordAnalogyComparison}
\vspace{-0.2in}
\end{table*}

%% file: texfiles/wiki-markup_applications.tex
As discussed earlier, the DAWT and other derived datasets that we have described in this paper have several applications for the NLP and IR communities. These include:

\subsection{Named Entity Recognition (NER)}
This task involves identifying an entity mention within a text and also generating candidate entities from the KB. For this, the Mention Occurrence, Entity Occurrence and the Mention To Entity Co-occurrence datasets described in Sections \ref{subsec:MentionOccur}, \ref{subsec:EntityOccur} and \ref{subsec:MentionEntityOccur} are extremely useful. 
For instance, the raw mention occurrence counts and probabilities can be stored in a dictionary and can be used to extract the mentions in a document. Furthermore, the mention occurrence count of a mention $M_{i}$  and its co-occurrence count with an entity $E_{j}$ can be used to calculate the prior probability of the mention mapping to that entity:
$${{count (M_{i} \rightarrow E_{j})} \over {count (M_{i})}}$$

This can be used to determine the candidate entities for a mention from our KB.

\subsection{Entity Disambiguation and Linking (EDL)}
This task involves linking the mention to its correct KB entity. For each mention, there may be several candidate entities with prior probabilities as calculated in NER. In addition, other features derived from these datasets can include entity co-occurrences, entity word2vec similarity and lexical similarity between the mention and the entity surface form. These features can be used to train a supervised learning algorithm to link the mention to the correct disambiguated entity among all the candidate entities as done in \cite{Bhargava:edl}. The approach used in \cite{Bhargava:edl} employs several such context dependent and independent features and has a precision of 63\%, recall of 87\% and an F-score of 73\%.

%% file: texfiles/wiki-markup_access.tex
The DAWT and derived datasets discussed in paper are available for download at this page:
\url{https://github.com/klout/opendata/tree/master/wiki_annotation}.
The DAWT dataset was generated using Wikipedia Data Dumps from January 20th 2017. Statistics regarding the data set are
shown in Table \ref{table:data_set_stats}.

%% file: texfiles/wiki-markup_related.tex
While a lot of works have focused on building and opening such datasets, very few have addressed all the challenges and constraints that we mentioned in Section \ref{section:problem_statement}. Spitkovsky and Chang \cite{SPITKOVSKY12.266} opened a cross-lingual dictionary (of English Wikipedia Articles) containing 175,100,788 mentions linking to 7,560,141 entities. This dataset, though extremely valuable, represents mention - entity mappings across a mixture of all languages which makes it harder to use for a specific language. In addition, this work used raw counts that, although useful, lack mention context (such as preceding and succeeding tokens etc.) which have a big impact while performing EDL. The Freebase annotations of the ClueWeb corpora dataset\footnote{http://lemurproject.org/clueweb09/FACC1/} dataset contains 647 million English web pages with an average of 13 entities annotated per document and 456 million documents having at least 1 entity annotated. It does not support multiple languages.

Another related technique for generating such dictionaries is Wikification \cite{medelyan2009mining, Milne:2008:LLW:1458082.1458150, cheng2013relational} where mentions in Wikipedia pages are linked to the disambiguated entities' Wikipedia pages. Such techniques rely on a local or global approach. 
A local approach involves linking observed entities using only their local context eg. by comparing the relatedness of candidate Wiki articles with the mentions \cite{cucerzan2007large, ferragina2010tagme, skaggs2014topic} while in global approach entities across the entire document are disambiguated together using document context, thus, ensuring consistency of entities across the document \cite{kulkarni2009collective, ratinov2011local}. 
Most recently Cai et al. \cite{Cai:2013:WVL:2541154.2505521} achieved 89.97\%
precision and 76.43\% recall, using an iterative algorithm that leverages link graph, link distributions, and a noun phrase extractor.

Although the problem of entity linking has been well studied for English, it has still not been explored for other languages. McNamee et al.  \cite{mcnamee2011cross} introduced the problem of cross-language entity linking.
The main challenge here is that state-of-the-art part-of-speech taggers perform much better on English than on other languages. In addition, both Wikipedia and Freebase have significantly higher quality and coverage of English compared to any other language.











%% file: texfiles/wiki-markup_conclusion.tex
In this work, we opened up the DAWT dataset - Densely Annotated Wikipedia Texts across multiple languages. The annotations include labeled text mentions mapping to entities (represented by their Freebase machine ids) as well as the type of the entity. The data set contains total of $13.6M$ articles, $5.0B$ tokens, $13.8M$ mention entity co-occurrences.
DAWT contains 4.8 times more anchor text to entity links than originally present in the Wikipedia markup. Moreover, it spans several languages including English, Spanish, Italian, German, French and Arabic. We also presented the methodology used to generate the dataset which enriched Wikipedia markup in order to increase number of links. In addition to the main dataset, we opened up several derived datasets for mention occurrence counts, entity occurrence counts, mention entity co-occurrence counts, entity word2vec as well as mappings between Freebase ids and Wikidata item ids. We also discussed two applications of these datasets and hope that opening them up would prove useful for the NLPand IR communities as well as facilitate multi-lingual research.

In the future, we plan to improve the algorithm that we used for generating DAWT. Also, we plan to migrate from using Freebase ids in our KB to Wikidata item ids.

%% file: texfiles/wiki-markup_appendix.tex
\section{Sample annotated Wikipedia texts from DAWT}

\begin{figure*}[t]

    \begin{subfigure}{0.9\textwidth}
        \centering
        \includegraphics[width=180mm]{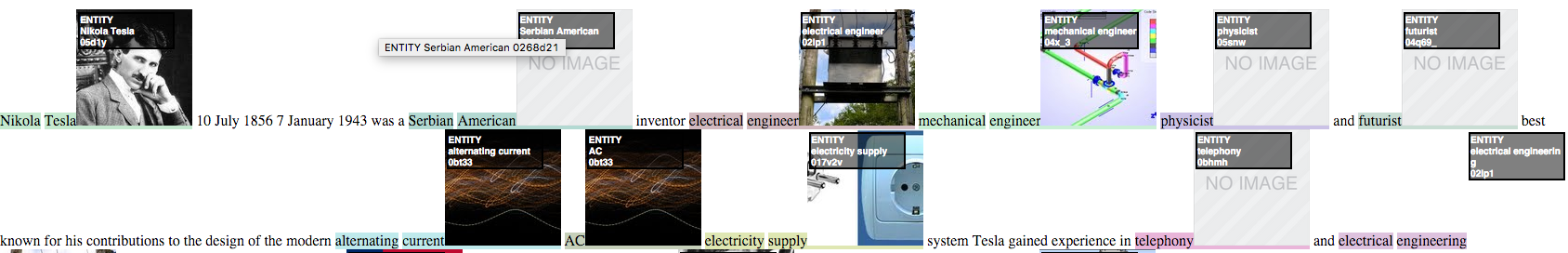}
        \caption{Annotated wikipedia article on Nikola Tesla (English)}
        \label{fig:nikola_tesla_en}
    \end{subfigure}

    \begin{subfigure}{0.9\textwidth}
        \centering
        \includegraphics[width=180mm]{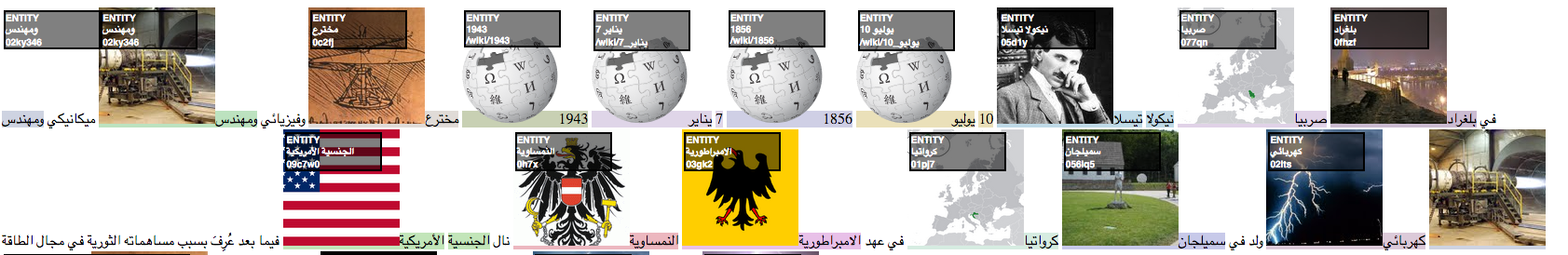}
        \caption{Annotated wikipedia article on Nikola Tesla (Arabic)}
        \label{fig:nikola_tesla_ar}
    \end{subfigure}

    \begin{subfigure}{0.9\textwidth}
        \centering
        \includegraphics[width=180mm]{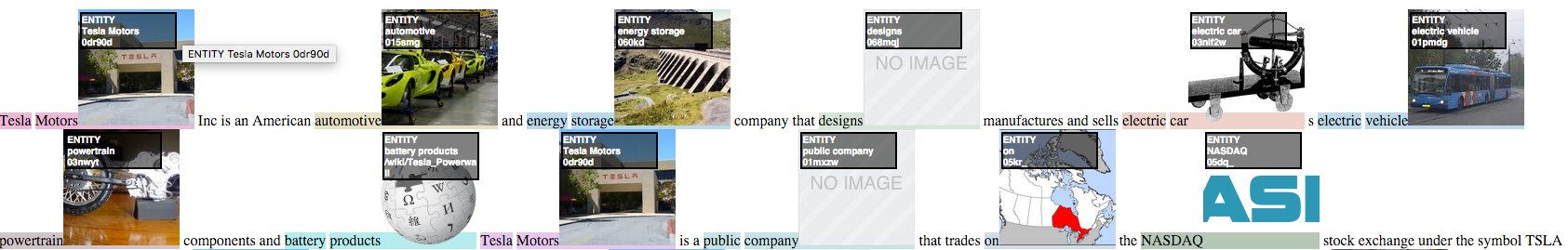}
        \caption{Annotated wikipedia article on Tesla Motors (English)}
        \label{fig:tesla_motors_en}
    \end{subfigure}
    \begin{subfigure}{0.9\textwidth}
        \centering
        \includegraphics[width=180mm]{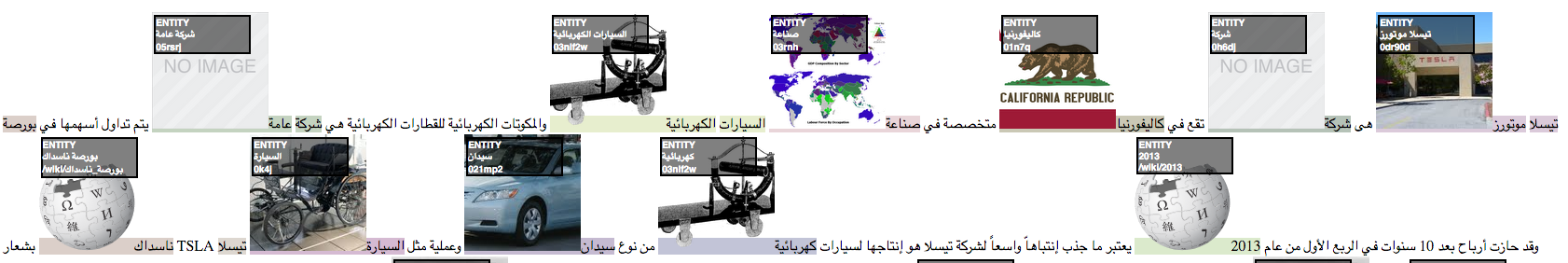}
        \caption{Annotated wikipedia article on Tesla Motors (Arabic)}
        \label{fig:tesla_motors_ar}
    \end{subfigure}
    \caption{Samples of extracted text across different Languages}
    \label{fig:annotatedsamples}
\end{figure*}

Figure \ref{fig:annotatedsamples} shows samples of the densely annotated Wikipedia pages for the entities \emph{Nikola Tesla} and \emph{Tesla Motors} across English and Arabic. 

